# Developers Struggle with Authentication in Blazor WebAssembly

Pascal Marc André
University of Bern, Switzerland
pascal.andre@students.unibe.ch

Quentin Stiévenart
Vrije Universiteit Brussel, Belgium
quentin.stievenart@vub.be

Mohammad Ghafari
TU Clausthal, Germany
mohammad.ghafari@tu-clausthal.de

*Abstract*—WebAssembly is a growing technology to build cross-platform applications. We aim to understand the security issues that developers encounter when adopting WebAssembly. We mined WebAssembly questions on Stack Overflow and identified 359 security-related posts. We classified these posts into 8 themes, reflecting developer intentions, and 19 topics, representing developer issues in this domain. We found that the most prevalent themes are related to bug fix support, requests for how to implement particular features, clarification questions, and setup or configuration issues. We noted that the topmost issues attribute to authentication in Blazor WebAssembly. We discuss six of them and provide our suggestions to clear these issues in practice.

## I. INTRODUCTION

WebAssembly (WASM) is a binary-format code that can be run in web browsers. It serves as a compilation target for other programming languages, and it aims to bring a near-native performance across platforms. For instance, Blazor WebAssembly is a framework that enables developers to build Web UIs by compiling C# code to WebAssembly.

Security has been a prime concern in the design of WebAssembly, and researchers have scrutinized the security of this technology since its inception [1]–[4]. However, there is no study yet regarding security concerns that developers face when they adopt WebAssembly. We believe such a study is essential to identify where our community can contribute and support this recent technology, which has received rapt attention in industry.

Therefore, we mined WebAssembly questions, answers, and comments that developers posted on the Stack Overflow website, and we investigated the following research questions:

**RQ1**: What are the intentions behind developer questions? We classified the posts into 8 *themes* with the top four being bug fix support, learning to implement a feature, clarification questions, and setup/configuration issues. We noted that questions related to bug fix support tend to be more frequently answered by the original authors, while questions related to best practice guidance and how-to instructions are least answered by their authors.

**RQ2**: What are the problem areas where developers need support? We assigned developer questions to 19 *topics* and found that the problem areas that developers seek help the most concern the Blazor framework, which appears in over 60% of all questions, and Authentication, which appears in half of the questions. The subsequent topics are led by "Azure", "User Token" and "HTTP Requests", which each have a share of roughly 10%. We discuss six most commonly faced issues, which are related to Authentication hassles in Blazor WebAssembly.

In summary, this study is a first attempt to draw our community's attention to the areas where WebAssembly technology may be improved especially to bootstrap the secure development of cross-platform applications in Blazor.

We share a replication package containing the full dataset, scripts, and analysis results to advocate future studies in this area.[1]

## II. RELATED WORK

The existing body of research is mainly concerned about the design of WebAssembly, which may have security consequences when compiling vulnerable C code [3], [5], [6], or when including sensitive functions such as means of evaluating JavaScript in the application [5], [7].

Researchers have conducted empirical studies on security concerns discussed on the Stack Overflow website, but none of them include WebAssembly. For instance, Meng et al. showed that in the Java ecosystem, the main concerns are related to APIs, libraries, cross-language data, and Java-specific practices [8]. Hazhirpasand et al. highlighted developer obstacles to adopting cryptography [9]. To the best of our knowledge, the focus of empirical studies for WebAssembly has been merely on collecting and analysing datasets of real-world WebAssembly binaries [1], [10]. We believe that uncovering the intentions behind developer issues in WebAssembly and the problem areas where developers struggle when they adopt this technology is also necessary to shed light on the areas where our community can support this growing technology.

## III. METHODOLOGY

We explain how we construct a dataset of security-related WebAssembly posts from the Stack Overflow website, and we describe our methodology to inspect these posts.

**The dataset**: We curate a dataset that comprises questions related to the security of WebAssembly, along with all of their answers and comments. To find the relevant questions, we rely on a recent list of 35 security-related keywords [11], which we expect to appear in most of such questions.

---
[1] https://figshare.com/articles/dataset/Dataset/19184795

| Security keywords |
|---|
| attack, exploit, risk, danger, threat, compromise, login, authentication, password, privilege, permission, encryption, leak, breach, injection, sanitize, overflow, bypass, malicious, defense, protect, sandbox, untrusted, trusted, trustworthy, secret, isolate, hazard, expose, signin, verification, certification |

We manually process the keywords to find all possible variations of them that help us to find as many Stack Overflow questions as possible. In effect, for each keyword we add, if available, its noun (singular and plural), verb (incl. present and past participle), adverb, adjective (several forms by adding suffixes such as "-able", "-ed" or "-ing") and other similar word variations we could find. We also include different variations of how a word may be written, for example, "login", "log-in", or "log in". In addition, the noun form of a word may have multiple variations too (e.g., "defend" has the "defense" and "defender" variations). We include all such variations, which eventually result in a total of 266 keywords.[2]

We query the Stack Overflow posts (title and body) using each of these keywords, combined with the "WebAssembly" keyword. We collected a total of 467 questions, along with all details of its answers and comments.

We then inspect all 467 questions manually and rate each with a score from 0 to 3 to measure how strongly they are related to WebAssembly security. We assign 0 to questions that were obviously irrelevant (e.g., post 62114179 asks for reasons other than security to use WebAssembly), and 1 to posts that were not clearly security-related (e.g., post 66923063 asks how to exclude a shared library when building a WebAssembly application). A rate of 2 is assigned to questions that have a weak link to WebAssembly security i.e., that do not have a direct security implication, but where a security concern is mentioned along the discussions (e.g., in post 66733106, a user asks how the JavaScript and WASM virtual machines work. However, during the discussion the author asks whether the virtual machines provide isolation). Finally, we assign 3 to questions that had a strong link to security, meaning its topic has a direct impact on protecting a system from threats (e.g., post 66873168 asks about best practices to implement user management modules).

108 (23%) of the questions are assigned level 0 or 1 and are therefore irrelevant to this study. In such questions, we see keywords such as "log in" that cause false positives, as they match questions related to logging data to the console. In the same vein, "private" matches questions related to the access modifier, and "sign" matches questions related to integer signedness.

The remaining 359 questions (77%) that are rated with level 2 or 3 are security-related WebAssembly questions and form our final dataset, which we simply call *dataset* in this paper.

[2]This strategy matches a few more questions compared to word stemming.

**Manual inspection**: We are interested to know what are the *themes* of each question, i.e., the intentions of the author for asking a question, and the *topics* of each question, i.e., the problem area where a developer needs help.

We do not have an initial list of themes that we can assign to the questions. Therefore, we resort to the *open card sorting* approach to identify the intention of each post. We pick the first question and record its theme as "Theme A". Then we go to the second question and identify its theme. If the identified theme corresponds to the one of "Theme A", we assign it to this theme, otherwise we create the new "Theme B" and assign the question to this theme. We repeat this process for all questions until each is assigned to at least one theme.

Each question can be assigned to several themes. We assign a scale from 0-3 to each question to indicate how strongly a theme was identified in the question. Level 1 indicates a weak link, whereas 3 indicates the main primary theme of a post.

We assign to each theme in our final list a title and a short description. To test our categorisation process and ensure that the results are reliable, two authors of this paper conducted a pilot study where they both individually and independently analysed the same 20 questions. We then shared and discussed our decisions. This helped us to verify our process and find any differences on how we might have evaluated a question to make sure that we get to the same conclusions and our process becomes more consistent to reduce inaccuracies in our findings. We resolved any conflict by a group discussion where each author provided a reasoning to convince the other.

We perform a similar pass (open-card sorting approach) over the dataset to identify question topics i.e., the areas to which a problem description relates the most (e.g., frameworks, services, compilers, infrastructure, software features, and technical concepts). For each post, we mark the presence of a topic (i.e., with a boolean value), and we assume that a post can have multiple topics. In the end, we identified a total of 19 different question topics.

## IV. RESULTS

Our final dataset includes 359 questions, 357 answers, and 1 097 comments that we inspected carefully. In total, 631 distinct users are involved in these activities (312 question authors, and 109 that provide an accepted answer). Each Stack Overflow question can be assigned one or more tags by the users of the website. In total 234 unique tags are assigned to the 359 questions a total of 1 348 times, which is an average of 3.79 tags per question. We noticed that a majority of tags are related to the Blazor framework. We also noted a massive increase in the number of questions in 2020, which may be attributed to the official release of Blazor WebAssembly.

We looked at the status of each post and found that 229 questions (64%) are marked as answered which means their issues have been resolved, whereas 130 questions (36%) still need to be addressed. We wondered why 130 questions were not resolved yet, and we found that 18 questions lack key details, and 28 questions were either too specific or too broad.

Notably, we could not associate any reason to 48 unanswered questions, which require a detailed investigation in future.

*A. RQ1: What are the intentions behind developer questions?*

We identified eight themes that reveal the types of questions that developers ask and the reasons for asking such questions. Table I presents these themes, their explanations, and the distribution of developer issues that relates to each theme.

Most questions are asked with the intent of obtaining support to fix a bug in the code as well as getting support on how to implement certain features. The occurrence of clarification questions, configuration issues, and best practice are also frequent. In contrast, questions that concern "unexpected results", "not supported" and "third-party bugs" are rare.

We note that the highest percentage of unanswered questions is in the "Best Practice Guidance" theme, where 40% are unresolved and thus tend to be more difficult to answer. The "How-To Instructions" theme is also above average, at 38%. In contrast, questions on installation and configuration problems are the most likely to be answered, with 20% still unanswered.

We observe that 20% of the questions are answered by the original author. In particular, installation and configuration problems as well as bug fix support questions tend to be more frequently answered by their original authors (resp. in 43% and 30% of the cases). Indeed, such questions can be solved through one's own efforts with debugging, research as well as trial and error. Best practice guidance (16%) and how-to instructions (15%) are clearly behind, with clarification questions being least answered by their authors with 11%. For these themes, solving the question is more difficult since the necessary experience cannot be acquired so easily, and the questions were in the first place asked to acquire more knowledge in the domain.

> **Developer Intents**
>
> The most questions are asked with the intent of obtaining support to fix a bug in the code as well as getting support on how to implement certain features. The highest percentage of unanswered questions is in the "Best Practice Guidance" theme. Installation and configuration problems as well as bug fix support questions tend to be more frequently answered by their original authors.

We also investigated whether themes vary in terms of the number of views, the number of responses, question scores and response times, but we did not find any significant difference.

*B. RQ2: What are the problem areas where developers need support?*

We investigated problem areas that developers seek help the most and identified 19 topics. Each topic had more than five occurrences in our dataset. Table II lists these topics, provides the description of each topic followed by an example post, and it presents the distribution of questions in each topic.

We noticed that Blazor WebAssembly is the main topic that developers ask about, as 61% of all questions in the dataset are about this topic. Very popular and far ahead of all others are authentication questions, encountered in 50% of all questions.

Each question may belong to more than one topic. We check how often certain pairs of topics appear together in a question, and we find that all questions about "Authentication" are highly related to the "Blazor" topic. The same is true for "Azure" and "Blazor". When developers ask questions about the theme of "How WebAssembly works", we find that most issues are about "How WebAssembly interacts with other programming languages". On the other hand, questions that belong to "Emscripten" topic, do not appear in those categorised as "Blazor" or "Authentication" topics. This topic mainly shows issues along with topics like "How WebAssembly works" or "How WebAssembly interacts with other programming languages".

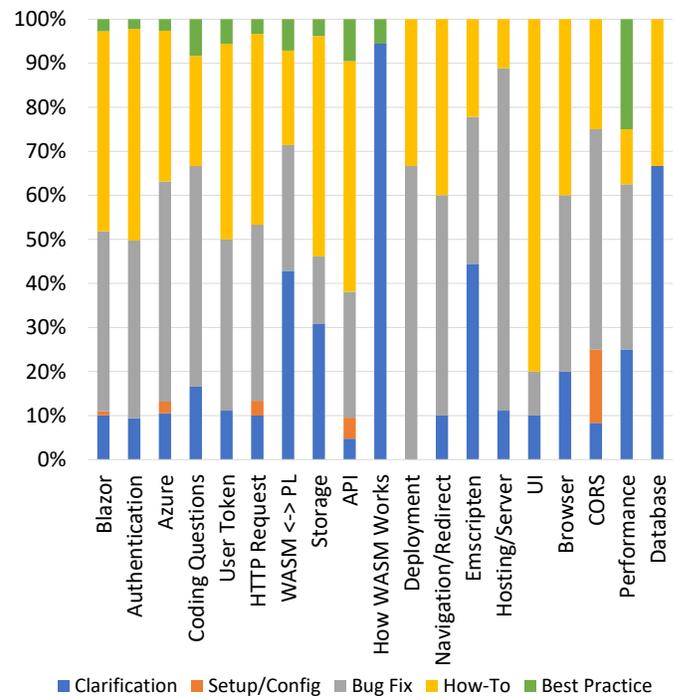

Fig. 1. The distribution of themes in topics

The distribution of question themes over each topic is illustrated in Figure 1. Questions covering issues on "How WebAssembly works" almost exclusively discuss "Clarification" themed questions and don't involve and bug fixes or how-to instructions. When it comes to security-related questions about user interfaces, developers mostly want to know how to implement particular features such as customising the default login form of a Blazor app or adjusting the layout depending on the currently signed-in user's role. "Bug Fix" themed questions have their highest share in the two topics related to the deployment and hosting of an application. Developers do not seem to ask many questions about this topic to clear their doubts as "Clarification"-themed questions score low numbers but they seem to have a relatively high ratio of bug-related issues. Finally, CORS related problems feature the highest share of setup and configuration issues.

TABLE I
THE EIGHT IDENTIFIED THEMES

| Id | Theme | Description | Questions |
|---|---|---|---|
| 1 | Bug Fix Support | Help on resolving a problem with a bug he cannot fix. | 181 (50%) |
| 2 | How-To Instructions | Specific instructions on how to implement a certain feature. | 168 (47%) |
| 3 | Clarifications | Conceptual clarifications on a given topic that they do not fully understand yet and need help with. | 99 (28%) |
| 4 | Setup/Config Issue | Help regarding an error that is related to an incorrect setup or configuration. | 79 (22%) |
| 5 | Best Practice | Best practice guidance on how to solve an issue or implement a certain feature. | 43 (12%) |
| 6 | Unexpected Results | Help on an issue with unexpected behaviour or error message that confuse them. | 8 ( 2%) |
| 7 | Not Supported | Help to implement a feature that is not supported by a tool, or to use a tool in a non-supported way. | 7 ( 2%) |
| 8 | Third-Party Bug | Help regarding an issue that is caused by a bug in a third-party software. | 4 ( 1%) |

TABLE II
THE 19 IDENTIFIED QUESTION TOPICS. THE EXAMPLE COLUMN CONTAINS A LINK TO A STACK OVERFLOW QUESTION WITHIN EACH TOPIC.

| Id | Topic | Description | Example | Questions |
|---|---|---|---|---|
| 1 | Blazor | Use of Microsoft Blazor with WASM to run the client-side C# code in the browser | 62700005 | 218 (61%) |
| 2 | Authentication | Authentication procedures | 66955557 | 181 (50%) |
| 3 | Azure | Azure services, including Azure Active Directory (AAD), Blob Storage, B2C, etc. | 65808332 | 38 (11%) |
| 4 | User Token | Create, send and manage tokens used for procedures regarding authentication | 62306681 | 36 (10%) |
| 5 | HTTP Requests | How to send HTTP requests, perform CRUD operations, set headers or send tokens | 63831943 | 30 ( 8%) |
| 6 | Storage | File-system and local storage operations, or to memory and cache related issues | 45535301 | 26 ( 7%) |
| 7 | API | APIs to perform CRUD operations, changing configuration options, managing endpoints, etc. | 63782180 | 21 ( 6%) |
| 8 | Navigation/Redirect | Navigation within an app and redirecting a user to a different location | 64947553 | 20 ( 6%) |
| 9 | Deployment of app | Building, publishing and deploying a new app or an already published or deployed app | 60932702 | 18 ( 5%) |
| 10 | How WASM works | How WASM works in general or to understand one of its specific features | 56506468 | 18 ( 5%) |
| 11 | WASM ↔ PL | Interactions between WebAssembly and other programming languages or tools | 50149603 | 14 ( 4%) |
| 12 | Coding Questions | Syntax errors or compilation issues | 65316643 | 12 ( 3%) |
| 13 | CORS | How to enable CORS, disable it, configure it, and fix errors caused by it) | 66960982 | 12 ( 3%) |
| 14 | User Interface | Adding elements to a UI, adjusting their behaviour or modifying template components | 65979821 | 10 ( 3%) |
| 15 | Emscripten | The Emscripten compiler that can compile in C and C++ source code to WebAssembly | 63952910 | 9 ( 3%) |
| 16 | Hosting/Server | Hosting an application, to its setup and configuration | 61736263 | 9 ( 3%) |
| 17 | Performance | Performance issues, such as understanding or improving the performance of a process | 64087065 | 8 ( 2%) |
| 18 | Database | Databases covering CRUD operations on structured data stored in a RDBMs like SQLite | 56500709 | 6 ( 2%) |
| 19 | Browser | Web browsers such as developer tools, or browser-specific behaviours | 56940990 | 5 ( 1%) |

We observed six issues that developers frequently encountered. These issues, listed below, mostly concern "Blazor" and "Authentication" indicating the areas where developers struggle the most.

- *How to implement third-party authentication?* More specifically, developers ask for code snippet to combine multiple social providers for authentication (e.g., post 63763449) or questions specific to JWT tokens (e.g., post 64888884). While the documentation is clear for combining multiple social providers[3], Blazor should provide an option during initial project setup to automatically add social providers to the project or have them included in an extended demo project for developers to see how they can implement it.
- *Why does authentication work locally in development mode but not after deployment?* The most common cause here is related to certificates, where developers are not aware of their existence (e.g., post 65823274), or did not set certificates correctly (e.g., post 65599957). A troubleshooting guide or checklist with the most common reasons for this kind of problem should be provided in the documentation.
- *Why user navigation does not work during authentication?* Issues vary here between failing to navigate to a specific page after login (e.g., post 64947553), failing redirections during login (e.g., post 62406340), or failing to navigate to the login page (e.g., post 65295133). Due to the high number of problems around navigation during authentication, a checklist or troubleshooting guide should also be provided here, in order to help developers identify the reasons of their issue.
- *How to customize the pre-existing UI elements of the login screen?* While unofficial tutorials exist for this issue, the official documentation section that covers this issue[4] is buried in a long page, difficult to find through search queries. We advise to move such documentation to a specific page. Moreover, providing a solution through specific guides labelled with a question such as "How to customize UI Blazor login page?" would provide further help. The demo project of the framework could link the login page to its documentation that explains on how to customize it.

[3]https://docs.microsoft.com/en-us/aspnet/core/security/authentication/social/?view=aspnetcore-6.0&tabs=visual-studio

[4]https://docs.microsoft.com/en-us/aspnet/core/security/authentication/scaffold-identity?view=aspnetcore-6.0&tabs=visual-studio#layout-and-authentication-flow-changes

- *How to assign user roles?* All such questions have been asked shortly after the official release of Blazor WASM. Similarly to the previous question, this topic is covered in the official documentation but difficult to find. Such issues should be featured in an extended demo application, and the documentation should contain a separate article that details how to implement it.
- *How to secure API endpoints to only give access to a group of selected or authenticated users?* These are usually developers that have a rough idea of what they want to achieve in terms of endpoint security, but do not possess enough information to implement it (e.g., 65365508). The documentation should be improved with a detailed starter guide leading new developers through an example on how to secure API endpoints to only give access to selected users. A demo application could moreover share advanced examples for CRUD operations.

We separately inspected posts that are not related to the Blazor technology and found that they mainly revolve around file system access, browser information access, memory leaks, and CORS issues. Nevertheless, such questions are very few at the moment and future studies are necessary when other framework and languages will become popular.

> **Topmost Issues**
>
> The main area where developers need support relates to Blazor WebAssembly and authentication. They frequently asked about implementing third-party authentication, authentication issues in deployment, user navigation and authentication, customization of UI in login pages, assigning roles, and access control to API endpoints.

## V. THREATS TO VALIDITY

The internal threats to the validity of this study stem from the manual inspection of developer posts and discussions. We resort to an open-card sorting approach to assign both developer intentions and question topics to the set of questions that we gathered. This could introduce bias in our results. To account for the fact that one question may be asked with more than one intention, or may fit more than one topic, we do not restrict each question to one intention and one topic, but rather combine multiple intentions with a scale of 1-3, and assume multiple topics for each question. We also conduct a pilot study with two authors of this paper to verify our process and resolve any inconsistencies.

The external threats to validity of this empirical study come from the way we build our dataset. We focus only on what developers discuss on the Stack Overflow website, which may bias our results towards certain types of questions. Nevertheless, Stack Overflow is one of the most popular platforms for asking programming-related questions, and we are not aware of any other forum where developers actively discuss WebAssembly questions. In addition, we rely on a set of security-related keywords to identify relevant posts. These keywords may not be comprehensive to find every single relevant post.

## VI. CONCLUSIONS

WebAssembly is a growing technology to build cross-platform applications. This work aims to draw our community's attention to the areas where this technology may be improved especially to bootstrap the secure development of cross-platform applications. We studied 359 security-related WebAssembly questions on Stack Overflow and found that in more than half of the posts, developers look for help on bug fixes and guidance on how to implement a particular feature. Other prevalent intentions when asking questions relate to clarification questions and setup or configuration issues. We identified 19 areas where developers seek help and observed that Blazor and authentication are by far the most widespread ones. We reported six most frequent issues that relate to authentication and provided actionable advice to address them.

We envision three main avenues for future work. Firstly, developer issues in this area may vary depending on the source language, the use of a specific framework, or even the target platform. Hence, a future investigation with respect to each of these aspects may yield new insights. Secondly, future research may build on this work and develop automated models that identify the themes and topics, as well as the sentiments of the questions. Finally, including other information sources such as WASM-specific developer forums may provide new insights.